# On Paradoxical Phenomena During Evaporation and Condensation between Two Parallel Plates

Gang Chen*

Department of Mechanical Engineering
Massachusetts Institute of Technology

**Abstract**

Kinetic theory has long predicted that temperature inversion may happen in the vapor-phase for evaporation and condensation between two parallel plates, i.e., the vapor temperature at the condensation interface is higher than that at the evaporation interface. However, past studies have neglected transport in the liquid phases, which usually determine the evaporation and condensation rates. This disconnect has limited the acceptance of the kinetic theory in practical heat transfer models. In this paper, we combine interfacial conditions for mass and heat fluxes with continuum descriptions in the bulk regions of the vapor and the liquid phases to obtain a complete picture for the classical problem of evaporation and condensation between two parallel plates. The criterion for temperature inversion is rederived analytically. We also prove that the temperature jump at each interface is in the same direction as externally applied temperature difference, i.e., liquid surface is at a higher temperature than its adjacent vapor on the evaporating interface and at a lower temperature than its adjacent vapor on the condensing interface. We explain the interfacial temperature jump and temperature inversion using the interfacial cooling and heating processes, and we predict that this process can lead to a vapor phase temperature much lower than the lowest wall temperatures and much higher than the highest wall temperature imposed. When the latent heat of evaporation is small, we found that evaporation can happen at the low temperature side while condensation occur at the high temperature side, opposing the temperature gradient.

* gchen2@mit.edu

## I. INTRODUCTION

Evaporation and condensation processes are ubiquitous in nature and industrial technologies [1–4]. Half a century ago, Pao [5] pointed out that the temperature gradient in the vapor region could be opposite to the externally imposed temperature difference for evaporation and condensation between two parallel plates. Subsequent studies based on solving the Boltzmann transport equation (BTE) [6–10] and molecular dynamics simulations [11–13] further confirm this prediction. Such inverted temperature profile in the vapor region is counter intuitive and its validity had since been questioned [2]. Koffman et al. [14], who also arrived at the temperature inversion results, wrote "this paradox is significant in that it casts a shadow of doubt on the fundamental theory." Aoki and Cercignani exclaimed [9] "A more spectacular result is the fact that in a two-plate experiment the temperature of the vapor at the hot wall can be below that at the cold wall …".

Previous kinetic theory-based treatments for the parallel-plate evaporation-condensation problem considered mostly the vapor-phase transport only, completely neglecting transport in the liquid films, despite that it is well-known in heat transfer that the liquid-layer thickness determines the mass flow rates. An exception is the work of Gatapova et al. [15] who considered the evaporating and condensing liquid layers in their treat of this problem. Their studies used linearized boundary conditions for temperature and pressure discontinuities obtained by Sone et al.. [16] and included heat conduction only in the vapor phase transport, neglecting the enthalpy carried by the evaporating mass flux, which could dominate the vapor phase transport. Molecular dynamics simulations [11–13] naturally include both the liquid and the vapor phases. However, the small simulation domain limits generality of the results. For the two parallel plates evaporation and condensation problem, the kinetic theory and the molecular dynamics simulations predict that at an evaporating interface, the liquid surface temperature is higher than that of the vapor phase at the outer edge of the Knudsen layer, while at the condensation interface, the vapor temperature at the outer edge of the Knudsen layer is higher than the liquid surface. The vapor-phase temperature is often inverted, i.e., the temperature at the condensation side is higher than at the evaporation side. So far, there exists no conclusive experiment on the temperature inversion for the two parallel-plate configurations. Experiments on single interface evaporation mostly reported an opposite sign of the interfacial temperature jump [2,17]. Hence, controversies with regards to if the inversion exists and the sign of interfacial temperature discontinuities remain unsolved.

It should be reminded that phase-change processes are fundamentally driven by both the temperature and the chemical potential differences. Recently, the author developed a set of expressions for the interfacial mass flux and heat flux [18] using diffusion-transmission boundary conditions [19], connecting the saturation vapor temperature and density on the liquid surface to that of the vapor-phase temperature and density at the outer edge of the Knudsen layer. The mass flux expression is identical to the classical Hertz-Knudsen-Schrage theory [20], but the additional interfacial heat flux expression enables one to calculate the temperature and density difference between the saturated vapor on the liquid surface and that at the outer edge of the Knudsen layer, i.e., the interfacial density and temperature discontinuities. Since these expressions are derived based on the diffusion approximation of the kinetic theory in the vapor phase, they can be consistently coupled to the continuum treatment on both the liquid and vapor phases. This treatment led to the insight that the vapor adjacent to an evaporating interface is cooled and that to a condensing interface heated[21], resulting in an intrinsic temperature drop from the liquid phase to the vapor phase in evaporation and an opposite trend in condensation, and temperature inversion even at a single interface. However, for a single interface, the reverse conduction of heat accompanying the inverted temperature profile can lead to a reversal of the temperature drop at the interface, a higher temperature in the vapor-phase compared to the liquid surface in evaporation, for example. These predictions reconcile the difference between theory and experiments on a single interface.

In this work, I will apply the same approach to treat the parallel-plate evaporation-condensation problem, including heat transfer in the evaporating and condensing liquid films. I will present a clear analytical derivation of the condition for the temperature inversion in the vapor phase that is consistent with results obtained previously from kinetic theory. I also prove analytically that the interfacial temperature jump is in the same direction as externally applied temperature difference for all situations. Both inverted and noninverted temperature profiles will be shown, depending on the latent heat of evaporation. Our simulation results show that the vapor temperature can be colder at the evaporation interface and hotter at the condensing interface than that of the values specified on both walls. Furthermore, we predict that when the latent heat is small, evaporation can happen on the lower temperature side and condensation on the high temperature side, completely opposite to the heat flow direction. I also assess the potential contribution of the vapor-phase temperature inversion to energy recovery in membrane distillation processes. Although the temperature inversion is appreciable, the low thermal conductivity of the vapor phase limits the potential of heat recuperation based on the inversed temperature profile. Our work explains the existing paradoxes and predicts unexpected phenomenon. The modeling framework enables one to explore the full parameter spaces for the problem of evaporation and condensation between two parallel plates, and design experiments to validate the predictions.

## II. MATHMATHICAL MODEL

Consider evaporation and condensation between two parallel plates as shown in Fig.1 and neglecting the gravity effect, kinetic theory leads to the following expression for the heat flux in the bulk vapor phase

$$q \approx c_p T m - k\frac{dT}{dz} \tag{1}$$

where $c_p$ is the constant pressure specific heat and k the thermal conductivity. The mass flux is defined as $m = \rho u$, with u the average drift velocity. This definition requires that [21]

$$\rho T = Constant, \tag{2}$$

i.e., the pressure in the vapor phase is uniform. Since q and m are constant in the vapor phase, the above expression also leads to the energy conservation equation

$$c_p \dot{m} \frac{dT}{dz} = k \frac{d^2T}{dz^2} \tag{3}$$

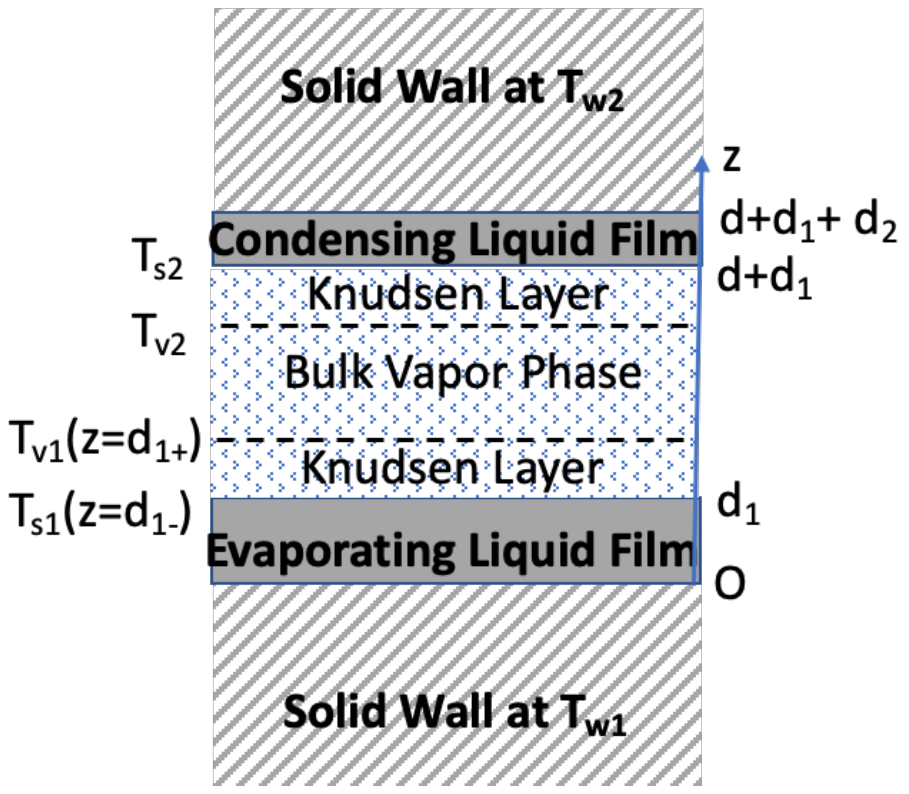

Figure 1 Modelled configuration and coordinate, the interface conditions neglect the Knudsen layer thickness.

On the evaporating liquid film side, the evaporation heat is supplied by the wall through the liquid layer. We neglect the impact of the evaporating mass flow on the temperature profile in the liquid, i.e., assume heat conduction only with a linearized temperature profile. Subtracting heat conducted to the interface by the latent heat of evaporation ($\Gamma$), we obtain the heat carried by saturated vapor at the evaporating interface as

$$q = k_w \frac{T_{w1} - T_{s1}}{d_1} - m\Gamma \tag{4}$$

where $T_{s1}$ is the liquid surface temperature which equals the saturation temperature of the vapor, $d_1$ the liquid layer thickness, and $k_w$ its thermal conductivity. We have neglected the temperature dependence of $k_w$ and $\Gamma$, which can be included if a more accurate model is needed. The first term in Eq.(4) is heat conduction from the wall to the interface. The second term reflects the fact that in the kinetic theory formulation of interface conditions, the outgoing molecules from the interface already overcome the interfacial latent heat barrier [18,22]. On the condensing liquid film side, a similar relation exists

$$q = k_w \frac{T_{s2} - T_{w2}}{d_2} - m\Gamma \tag{5}$$

Although rigorously speaking, $\Gamma$ should be the internal energy change rather than the enthalpy change during phase-change process,[21] no distinction is made in numerical calculations presented in this work. At the evaporating surface, $z=d_1$, we apply the classical Hertz-Knudsen-Schrage theory for the interfacial expression of the mass flux and the recently derived expression for the heat flux [18,20,21]

$$m = \frac{2\alpha_1}{2-\alpha_1}\sqrt{\frac{R}{2\pi}}\left[\rho_{s1}(T_{s1}){T_{s1}}^{1/2} - \rho_{v1}{T_{v1}}^{1/2}\right] \qquad (6)$$

$$q = \frac{4\alpha_1}{2-\alpha_1}R\sqrt{\frac{R}{2\pi}}\left[\rho_{s1}(T_{s1}){T_{s1}}^{3/2} - \rho_{v1}{T_{v1}}^{3/2}\right] \qquad (7a)$$

$$= 2RT_{v1}m + \frac{4\alpha_1}{2-\alpha_1}R\sqrt{\frac{R}{2\pi}}\rho_s\sqrt{T_s}(T_s - T_v) \qquad (7b)$$

where $\alpha_1$ is the accommodate coefficient, $\rho$ the density, R the ideal gas constant of the vapor, and T the temperature. The subscript "s" represents the properties of the saturated vapor phase on the liquid surface, and "v" the vapor phase properties at the outer edge of the Knudsen layer, which is of the order of a few mean free path lengths [23,24]. This thickness is neglected and hence "v" can be considered properties of the vapor phase immediately outside the liquid surface, as marked in Fig.1, i.e., "v" is at $z=d_{1+}$ while "s1" is at $z=d_{1-}$ Similarly, at the condensing interface, $z=d_1+d_v$, we have

$$m = \frac{2\alpha_2}{2-\alpha_2}\sqrt{\frac{R}{2\pi}}\left[\rho_{v2}{T_{v2}}^{1/2} - \rho_{s2}(T_{s2}){T_{s2}}^{1/2}\right] \qquad (8)$$

$$q = \frac{4\alpha_2}{2-\alpha_2}R\sqrt{\frac{R}{2\pi}}\left[\rho_{v2}{T_{v2}}^{3/2} - \rho_{s2}(T_{s2}){T_{s2}}^{3/2}\right] \qquad (9)$$

$\alpha_2$ is the accommodate coefficient at the condensing interface, which will be taken as identical to $\alpha_1$ in this work. Solution to Eq. (3) leads to

$$T(z) = \frac{T_{v2}\left[exp\left(c_p m(z-d_1)/k\right)-1\right]+T_{v1}\left[exp\left(c_p m d_v/k\right)-exp\left(c_p m(z-d_1)/k\right)\right]}{exp\left(c_p \dot{m} d_v/k\right)-1} \qquad (10)$$

Using the above equation, the heat transported in the vapor phase can be calculated from Eq. (1) as,

$$q = c_p T_{v1} m - \frac{m c_p (T_{v2}-T_{v1})}{exp\left(c_p m d_v/k\right)-1} = c_p T_{v2} m - \frac{m c_p (T_{v2}-T_{v1}) exp\left(c_p m d_v/k\right)}{exp\left(c_p m d_v/k\right)-1} \qquad (11)$$

where the first term represents convection and second term is due to conduction. The first term is always positive if the evaporation process happens at the interface (m>0) and negative for condensation (m<0). The second term, however, can be either positive or negative, depending on the sign of $T_{v2}$-$T_{v1}$. The expression itself, however, can be applied to both evaporation and condensation processes.

In the above equations, $T_{s1}$, $T_{v1}$, $\rho_{v1}$, $T_{s2}$, $T_{v2}$, $\rho_{v2}$, m, q are unknowns to be determined for given $T_{w1}$, $T_{w2}$, $d_1$, $d_v$, $d_2$ and other liquid properties. At the evaporation interface, equating (1), (4), and (7) leads to two independent equations and similarly at the condensation surface. The mass flux expressions (6) and (8), together with (2) and (11) give four other equations. These eight equations need to be solved concurrently. Numerical solution is further complicated by the nonlinear relationship between the saturate temperature and density. In this paper, we used the following correlation for saturated water vapor,[25]

$$\rho_s = 5.018 + 0.32321 \times t_s + 8.1847 \times 10^{-3} {t_s}^2 + 3.1243 \times 10^{-4} {t_s}^3 \tag{12}$$

where $t_s=T_s-273$ (°C). This fit works well for $t_s$ in [0, 40] °C. We also study other hypothetical fluids using Clausius-Clapeyron equation with a constant latent heat

$$\rho_s = \frac{\rho_r T_r}{T_s} exp\left[-\frac{\Gamma}{R}\left(\frac{1}{T_s}-\frac{1}{T_r}\right)\right] \tag{13}$$

where $T_r$ and $\rho_r$ are temperature and density at a reference point, which we take to be at $T_{w1}$. The nonlinear nature of the interfacial expression and the saturation properties make solving the coupled equations quite challenging. We reduce the eight coupled equations into two, expression all quantities using $T_{s1}$ and $T_{v2}$ as unknowns. For example, equating (4) and (5) allows expressing $T_{s2}$ in terms of $T_{s1}$. We can also prove the following relation holds true based on Eqs. (6)-(9) (assuming $\alpha_1=\alpha_2=\alpha$), which allows us to express $T_{v1}$ in terms of $T_{v2}$, $T_{s1}$, and $T_{s2}$.

$$\frac{\rho_{s1}(T_{s1}){T_{s1}}^{3/2}+\rho_{s2}(T_{s2}){T_{s2}}^{\frac{3}{2}}}{\rho_{s1}(T_{s1}){T_{s1}}^{1/2}+\rho_{s2}(T_{s2}){T_{s2}}^{\frac{1}{2}}} = \sqrt{T_{v1}T_{v2}} \tag{14}$$

The final two equations we solved are based on equating mass flux expressions (6) and (8), and heat flux expressions (4) and (11). Such two coupled equations can be readily solved with Matlab.

Since some of obtained temperature profiles presented later are counter intuitive, we will also calculate entropy generation to make sure that the transport process discussed does not violate the second law of thermodynamics. We had shown that the entropy generation in the vapor phase and the liquid phase are always positive under the given governing equations [21]. We calculate the entropy generation across the interfacial region, i.e., the Knudsen layer, using the expression established by Bedeaux et al. [26], recasting it into our notations

$$\sigma_i = -\frac{q(T_v-T_s)}{T_v T_s} - m\left[\frac{c_p(T_v-T_s)}{T_s} - \left(c_p ln\frac{T_v}{T_s} - R ln\frac{P_v}{T_s}\right)\right] \tag{15}$$

The above relation can be applied to both interfaces, using the corresponding quantities at each interface.

## III. RESULTS AND DISCUSSION

Figure 2(a) shows a calculated temperature profile for the specific conditions given in the figure, using properties close to that of water $\Gamma$=2.45 MJ/kg, $c_p$=1800 J/kg-K, $k_w$=0.6 W/m-K, $k_v$=0.02 W/m-K. We clearly see that the temperature distribution of the vapor phase is opposite to that of the liquid phases, and opposite to that of the externally applied temperature difference, i.e., inverted with a higher vapor temperature at the condensing interface higher than the evaporating one. In addition, the temperature discontinuity at each interface follows the flow direction: the liquid surface temperature at the evaporating interface is higher than the vapor phase, while the vapor phase temperature is higher than the liquid condensing liquid surface. Also shown in the figure is the vapor density distribution, with the density values at the liquid interface plotted artificially into the liquid region. The figure shows that the density drops monotonically from the evaporating side to the condensing side, consistent with the idea that chemical potential is the major driving force for mass transfer since density is related to the chemical potential. The entropy generation is always positive, as shown in Fig.2(b). In Fig.2(c), we show how the temperature distributions change with the changing liquid layer and the vapor gap thicknesses. As the liquid layer becomes thinner, the temperature drop at the interface increases because of increased evaporating flux. The vapor temperature can be significantly lower than the hot wall temperature on the evaporating side and much higher than the cold wall temperature on the condensing surface. This interfacial effect arises from the mismatch of the enthalpy at the interface $2RT_{v1}$ in Eq.7(a) and $2.5RT_{v1}$ in Eq. (1), taking $c_v$=5R/2 for monatomic ideal gas.[21] One can also understand this cooling effect as due to the sudden density drop at the interface. As the vapor gap $d_v$ becomes larger, the reverse heat conduction reduces, and hence the vapor temperature at the evaporating surface is even lower. On the condensing interface, the same enthalpy difference means vapor is heated up. The cooling at the evaporating interface and heating at the condensing interface leads to the temperature inversion predicted in the past. In Fig.2(d), temperature profiles are shown for different $d_1$ and $d_2$ values. The largest interfacial cooling was reached when the two sides are of equal thickness. Otherwise, the thicker liquid layers determine the mass flow rate, diminishes the temperature inversion. We notice that a thinner condensation film is more effective than a thinner evaporating film in creating lowest vapor temperature at the evaporating interface.

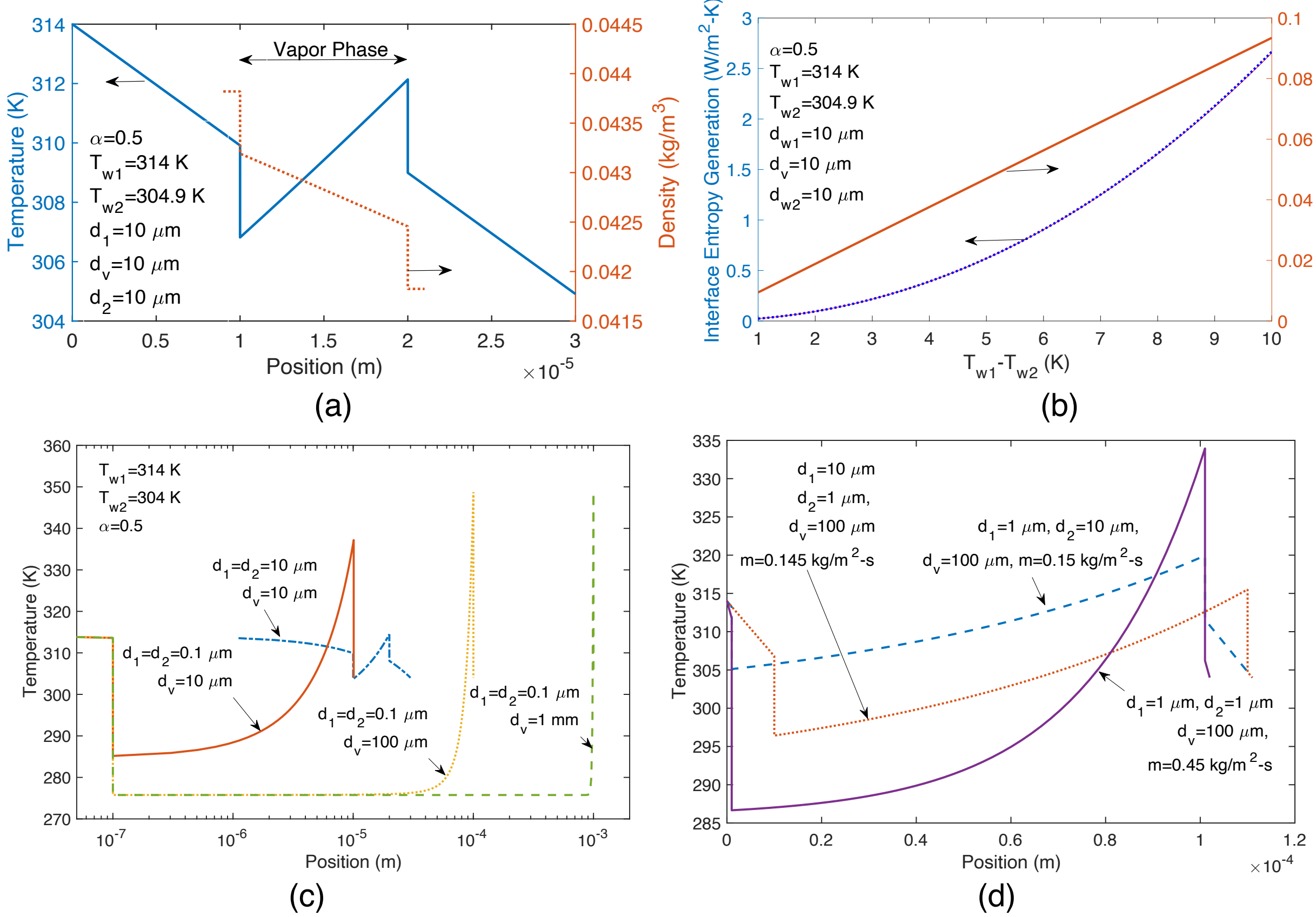

Figure 2. (a) Temperature and vapor density profile, (b) entropy generation at interface vs. temperature difference, and (c) temperature profile at different $d_1$, $d_2$, and $d_v$, (d) temperature profile when liquid films are of unequal thickness.

Based on the solution of the BTE, criteria for temperature inversion had been give before as

$$\beta = \frac{T}{\rho}\frac{d\rho}{dT} = \frac{\Gamma}{RT_o} - 1 \geq \beta_c \tag{16}$$

with different $\beta_c$ values, $\beta_c$=3.5 in Ref.[5], $\beta_c$=3.75 in Ref. [27], and 3.77 in Ref. [16]. In Eq. (16), the second equality is from the Clausius-Clapeyron equation. $T_o$ in Eq. (16) is taken as an average temperature. Here, we will provide a different analytical proof of Eq.(16), together with other key analytical results. For simplicity, we take $c_p$=5R/2, which was implicit in all past kinetic theories for monoatomic gas.

Summing up Eqs. (6) and (8), and (7) and (9), and using Taylor series to expand $\rho_{s1}$, $T_{s1}$ to $T_{s2}$, we have

$$2m \approx K\left\{\frac{\rho_{s2}}{T_{s2}^{1/2}}\left[\left(\frac{T}{\rho}\frac{d\rho}{dT}\right)_{T_{s2}} + \frac{1}{2}\right]\Delta T_s - \rho_{v1}T_{v1}\left(\frac{1}{T_{v1}^{1/2}} - \frac{1}{T_{v2}^{1/2}}\right)\right\} \tag{17}$$

$$2q \approx 2RK\left\{\rho_{s2}T_{s2}^{1/2}\left[\left(\frac{T}{\rho}\frac{d\rho}{dT}\right)_{T_{s2}} + \frac{3}{2}\right]\Delta T_s - \rho_{v1}T_{v1}\left(T_{v1}^{1/2} - T_{v2}^{1/2}\right)\right\} \tag{18}$$

where $\Delta T_s = T_{s1} - T_{s2}$ and $K=\frac{2\alpha_1}{2-\alpha_1}\sqrt{\frac{R}{2\pi}}$. Combining Eqs.(17)&(18) with (11), we can arrive at

$$\left[\beta(T_{s2})\left(1 - \frac{5}{4}\frac{T_{v2}}{T_{s2}}\right) + \left(\frac{3}{2} - \frac{5}{8}\frac{T_{v2}}{T_{s2}}\right)\right]\rho_{s2}T_{s2}^{\frac{1}{2}}\Delta T_s$$
$$= \rho_v T_v\left(\sqrt{T_{v1}} - \sqrt{T_{v2}}\right)\left[1 + \frac{5}{4}\sqrt{\frac{T_{v2}}{T_{v1}}} + \frac{5m\left(\sqrt{T_{v1}}+\sqrt{T_{v2}}\right)}{2K\rho_v T_v}\frac{exp(c_p m d_v/k)}{exp(c_p m d_v/k)-1}\right] \tag{19}$$

We can also further expand $T_{v2}$ around $T_{s2}$ and neglect the higher order terms, the above equation then leads to

$$\frac{1}{4}\left[\beta(T_{s2}) - \frac{7}{2}\right]\rho_{s2}T_{s2}^{\frac{1}{2}}\Delta T_s = \rho_{v1}T_{v1}\left(\sqrt{T_{v2}} - \sqrt{T_{v1}}\right)\left[1 + \frac{5}{4}\sqrt{\frac{T_{v2}}{T_{v1}}} + \frac{5m\left(\sqrt{T_{v1}}+\sqrt{T_{v2}}\right)}{2K\rho_{v1}T_{v1}}\frac{exp(c_p m d_v/k)}{exp(c_p m d_v/k)-1}\right] \tag{20}$$

Since the mass flow rate is positive, as long as $\Delta T_s = T_{s1} - T_{s2}$ is larger than zero and the left hand side is larger than zero as long as $\beta(T_{s2}) = \left(\frac{T}{\rho}\frac{d\rho}{dT}\right)_{T_{s2}} \geq \frac{7}{2}$, we have $T_{v2} \geq T_{v1}$, i.e., the vapor phase temperature is inverted. This means $\beta_c = 3.5$, identical to the result of Pao[5]. The approximations we made in equating $T_{v2}=T_{s2}$ from Eq. (19) to (20) also explains the small variations of $\beta_c$ arrived in the past literature. For water, $\beta \approx 16.6$ according to Eq.(16), the temperature in the vapor phase for water is always inverted.

The temperature inversion in the vapor phase means reverse heat conduction happens. For evaporation at a single interface, such reverse heat conduction can lead to a sign change in $T_{s1}$-$T_{v1}$, despite the intrinsic interfacial cooling effect gives a positive $T_{s1}$-$T_{v1}$ [21] For the coupled evaporation-condensation problem studied here, we can also prove that $T_{s1}$-$T_{v1}$ is always positive. To see this, we rewrite Eq. (14) as

$$T_{s1} - T_{v1} = T_{v1}\left(\sqrt{\frac{T_{v2}}{T_{v1}}} - 1\right) + \frac{\rho_{s2}(T_{s2}) T_{s2}^{\frac{1}{2}} \Delta T_s}{\rho_{s1}(T_{s1}) T_{s1}^{\frac{1}{2}} + \rho_{s2}(T_{s2}) T_{s2}^{\frac{1}{2}}} \tag{21}$$

The right-hand side is always positive as long as the temperature is inverted, hence, the temperature jump $T_{s1} - T_{v1}$ is positive. One can similarly proof that $T_{v2} - T_{s2}$ is positive on the condensation side. It is also easy to see that from Eq. (18) that $q \geq 0$. This is also different from the case of evaporation at a single interface [21], for which it is possible to have $q \leq 0$ when the reverse heat conduction is very large.

In the opposite case, when $\beta \geq \beta_c$, the temperature is not inverted, i.e., $T_{v1} \geq T_{v2}$. In this case, we can also prove $T_{s1}$-$T_{v1}$ is larger than zero. This can be easily seen by setting the interfacial heat flux equaling the bulk heat flux, which leads to

$$\frac{4\alpha_1}{2-\alpha_1} R \sqrt{\frac{R}{2\pi}} \rho_{s1} \sqrt{T_{s1}} (T_{s1} - T_{v1}) = \frac{1}{2} R T_{v1} m + \frac{m c_p (T_{v1} - T_{v2})}{exp(c_p m d_v / k) - 1} \tag{22}$$

Along as m is positive, the right-hand side is larger than zero, and hence the temperature drops from the liquid surface to the vapor side for the evaporating surface. Similarly, we can prove $T_{v2} \geq T_{s2}$.

As we will see later, there are special situations when m itself can be negative, i.e., evaporation happens from the lower temperature side. In this case, $T_{s1} \geq T_{v1}$ still holds true. This can be proven by (1) first rewriting Eq. (17) and (18) in terms of $\beta(T_{s1})$, i.e., using Taylor expansion to express saturation properties at $T_{s1}$, (2) eliminating the $-\rho_{v1} T_{v1}\left(T_{v1}^{1/2} - T_{v2}^{1/2}\right)$ term, and (3) utilizing Eq. (14), to arrive at

$$q \geq RK \rho_{s1} T_{s1}^{1/2} \Delta T_s \left\{ \left[\beta(T_{s1}) + \frac{3}{2}\right] - \frac{1 + \left(\frac{\rho_{s2}}{\rho_{s1}}\right)\left(\frac{T_{s2}}{T_{s1}}\right)^{3/2}}{1 + \left(\frac{\rho_{s2}}{\rho_{s1}}\right)\left(\frac{T_{s2}}{T_{s1}}\right)^{1/2}} \left[\beta + \frac{1}{2}\right] \right\} \tag{23}$$

We can easily see that right-hand side is positive. Hence, $q \geq 0$. Equation 7(b) then requires that $T_{s1} \geq T_{v1}$. Hence, in all scenarios, the sign of the interfacial temperature jump is the same as in the externally applied wall temperature difference direction.

Since we had assumed $\Delta T_s = T_{s1} - T_{s2}$ is always positive, it is useful to provide a proof too. We can take the two liquid surfaces at the liquid-vapor interfaces to enclose the vapor phase as the control volume, and apply the second law to such a control volume, which leads to the entropy generation

$$\sigma = q_w \left(\frac{1}{T_{s2}} - \frac{1}{T_{s1}}\right) \geq 0 \tag{24}$$

where $q_w$ is the conductive heat flux in the liquid film, i.e., the first term in Eqs.(4) and (5), which are equal to each other. The above relation ensures $T_{s1} - T_{s2} \geq 0$.

To study the temperature inversion criterion numerically, we choose an artificial liquid obeying the Clausius-Clapeyron relation, i.e., Eq. (13) or equivalently Eq. (16). Figure 3(a) shows that when β is less than $\beta_c$=3.5, the temperature inversion does not happen, as predicted. For all

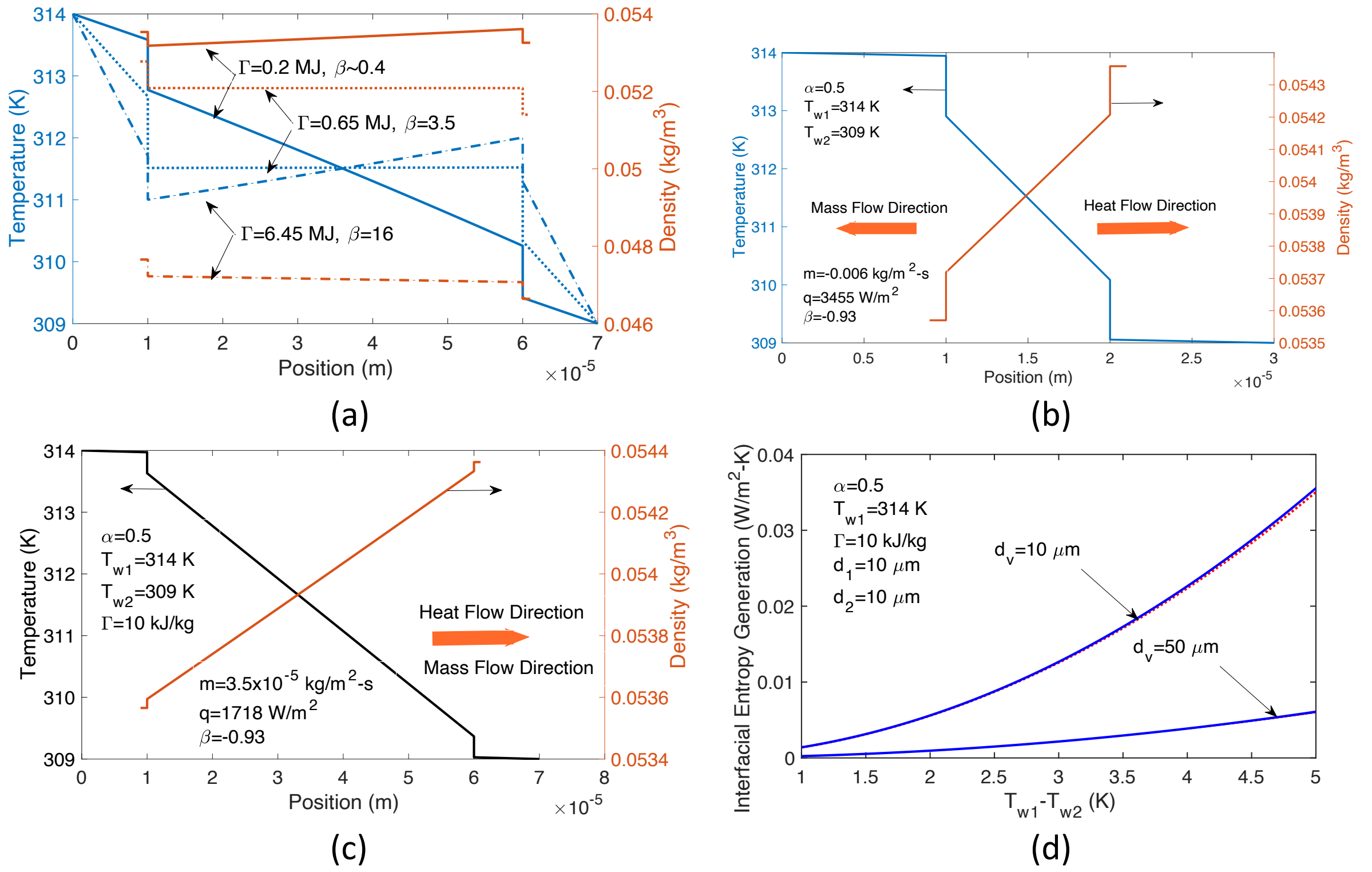

Figure 3 (a) Comparison of temperature and density distributions for different β values, showing that the temperature gradient changes sign at $\beta_c$=3.5. (b) Reverse evaporation/condensation for when $\beta \leq -0.5$, (c) when $\beta = -0.494$, the heat and mass flow directions are consistent, even though the density profile implies a negative mass flux, (d) interfacial entropy generation vs. temperature differences ((b)&(c) are special cases when $T_{w1}$-$T_{w2}$=5 K), showing positive entropy generation for the unconventional cases shown in (b)&(c).

cases, the pressure in the vapor phase is constant per assumption made in Eq.(2), and the temperature and density are inverse to each other, as Eq.(2) dictates. For these cases, we have checked that interfacial entropy generation is positive and the mass flux is positive, i.e., evaporation happens on the hot wall, and condensation on the cold wall.

A more spectacular simulation is shown in Fig.3(b), obtained using a much smaller latent heat of $\Gamma$=10 kJ/kg. For the specific conditions given, we found that the mass flow rate is negative, while the heat flux is positive. We have checked that the entropy generation values at both interfaces are positive, and hence the second law of thermodynamics is obeyed. This means that evaporation happens at the lower temperature side, condensing on the higher temperature side. This is completely against intuition.

We can understand this phenomenon from Eq.(17). When $\beta \leq -0.5$, the square bracket term in Eq. (17) is negative. Although the second term is positive in this case, it may not be large enough to overcome the first term. One example is shown in Fig.3(b). When $\beta$=-0.93, m is negative for the studied condition. In this case, heat conducts through the vapor region to the lower temperature side, leading to evaporation on this side. If $\beta$ is positive, density increases with increasing temperature. For an evaporating surface, temperature decreases and hence density decreases, as we see in Fig.2. However, when $\beta \leq -0.5$, the Clausius-Clapeyron equation dictates that density increases when temperature decreases at interfaces, as consistently shown in Figs.3(b) and 3(c). The reverse density profile, including interfacial density jump, leads to the mass transfer direction opposite to that of the heat transfer direction since density is measure of the chemical potential, which is the main driving force for mass transfer. In Ref. 21 we discussed the similarity and difference of the interfacial cooling effect with the Joule-Thomson effect and other known effects.

Figure 3(c) shows an even more subtle case. The only change from that of Fig.3(b) is a wider gap, $d_v$, from 10 μm to 50 μm. The mass flux is the same with the heat flux direction, although the density profile is similar to that of Fig.3(b). In this case, the temperature gradient driven mass transfer overwhelms density gradient driven mass transfer. Such a positive mass flux is possible since the Hertz-Knudsen-Schrage expression for the interfacial mass flux shows that it was the difference between $\rho T^{1/2}$, with $T^{1/2}$ comes from the molecular velocity, that determines the net mass flux. As $d_v$ becomes larger, less heat can be supplied to the cold side, causing evaporating from the hot side.

The inverted temperature profile in the vapor phase raises interesting question on if this phenomenon could be explored for improving the efficiency of some industrial process such as membrane distillation and multi-stage vapor-gap desalination[28–30], which relies on evaporation from one side and condensation on the other side. The recovered heat is the second part of Eq. (11). We can compare this term with the total heat supplied, and calculate the heat recovery ratio defined as

$$\gamma = \left[\frac{c_p(T_{v2}-T_{v1})\dot{m}}{exp(c_p\dot{m}d_v/k)-1}\right] \Big/ \left[\frac{k_w(T_{w1}-T_{s1})}{d_1}\right] \tag{21}$$

Figure 4 shows the calculated heat recovery ratio for some cases. Due to the low thermal conductivity of the vapor-phase, the heat recovery ratio is typically small.

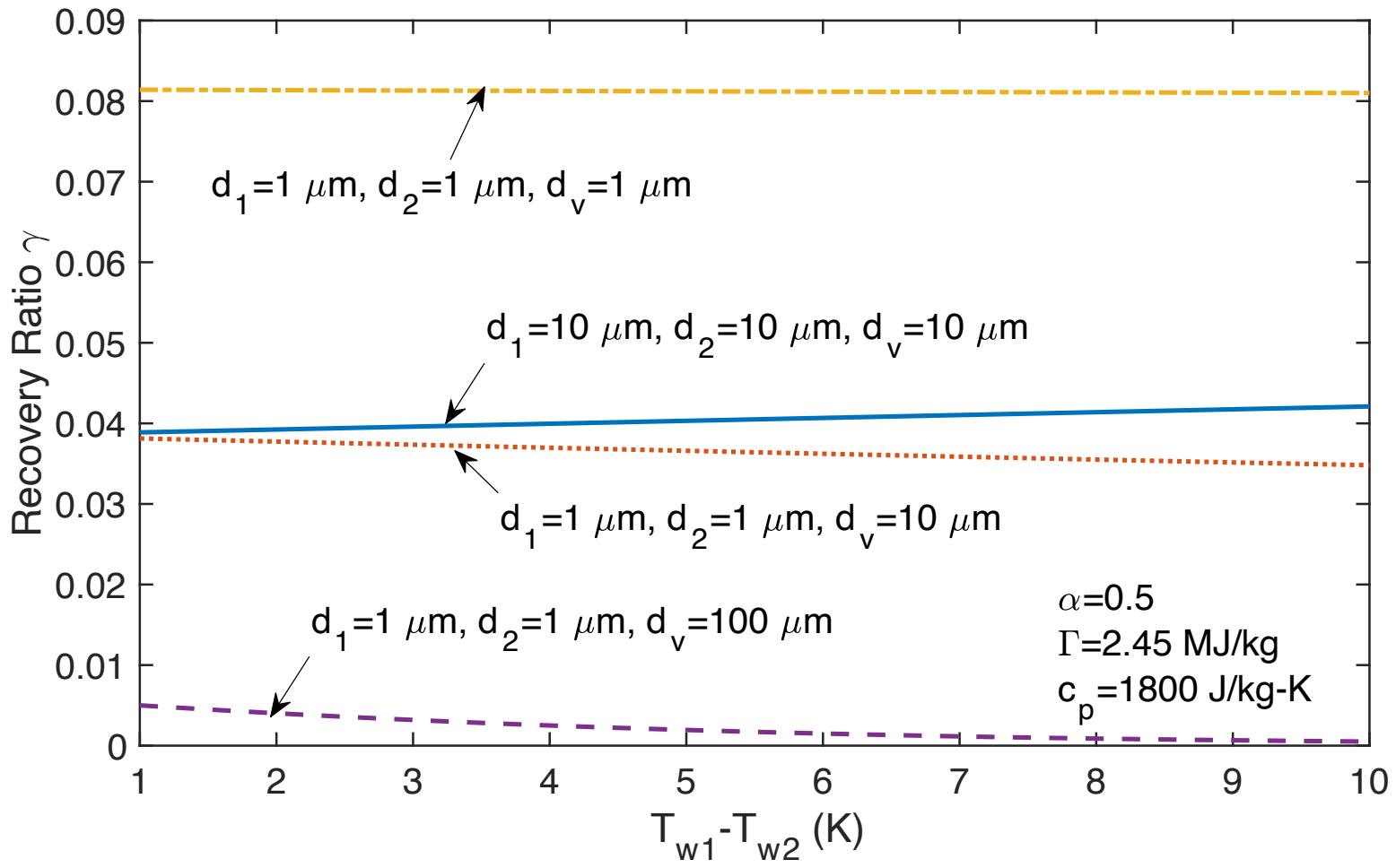

Figure 4 Heat recovery ratio due to inverted temperature profile

## IV. SUMMARY

In summary, we have shown rich behavior of evaporation and condensation between two parallel plates beyond what are already known as paradoxical. Past studies predicted the phenomenon of temperature inversion. However, difficulties in solving BTE had prevented a full exploration of the coupling between the liquid-phase and the vapor transport. We employ the recently developed interfacial heat flux expression, together with the classical Hertz-Knudsen-Schrage formula for mass transfer, to couple the transport in the liquid and the vapor phases. Our studies lead to both new physical insights and predictions. Consistent with previous predictions, an inverted temperature profile happens when $\beta = \frac{T}{\rho}\frac{d\rho}{dT}$ is larger than $\beta_c$=3.5. We further show that the vapor phase temperature can be lower even the cold wall temperature at the evaporating interface and higher than the hot wall temperature at the condensing interface. We interpret these phenomena as arising from the intrinsic cooling effect at an evaporating interface and heating effect at a condensing interface, caused by the mismatch in the enthalpy at the evaporating interface and the bulk vapor region, and accompanying sudden expansion of the vapor at both interfaces. We proof rigorously the temperature inversion criterion $\beta \geq \beta_c = 3.5$. We also prove that for all cases, the interfacial temperature jump is along the same direction as the external temperature difference, despite the existence of reverse heat conduction. In the range of $-1 \leq \beta \leq -0.5$, it is possible to have mass evaporating at the cold side and condensing at the hot side, opposite to the heat transfer direction. We evaluate the potential of exploiting the inverted temperature profile for heat recovery in membrane desalination, but found that the heat recovery fraction is not large. The framework we establish here will be useful for modeling heat and mass transfer in membrane distillation, desalination, as well as phase change heat transfer systems such as heat pipes and thermal ground planes.

**Acknowledgment.** This work is supported by MIT. I would like to thank Mr. Carlos Daniel Diaz Marin at MIT, Professors Hadi Ghasemi at University of Houston and Yanying Zhu at UCSB for their helpful comments on the manuscript.

**COI:** The author has not conflict of interests to disclose.

**Author's contribution:** conceptualization, formal analysis, methodology, software, visualization, writing.